# Efficient ion acceleration driven by a Laguerre Gaussian laser in near-critical-density plasma

Jia-Xiang Gao[1], Meng Liu[2†], and Wei-Min Wang[1,3,4††]

[1]*Department of Physics and Beijing Key Laboratory of Opto-electronic Functional Materials and Micro-nano Devices, Renmin University of China, Beijing 100872, China;*
[2]*Department of Mathematics and Physics, North China Electric Power University, Baoding, Hebei 071003, China;*
[3]*Key Laboratory of Quantum State Construction and Manipulation (Ministry of Education), Renmin University of China, Beijing, 100872, China;*
[4]*IFSA Collaborative Innovation Center, Shanghai Jiao Tong University, Shanghai 200240, China.*

Laser-driven ion accelerators have the advantages of compact size, high density, and short bunch duration over conventional accelerators. Nevertheless, it is still challenging to generate ion beams with quasi-monoenergetic peak and low divergence in the experiments with the current ultra-high intensity laser and thin target technologies. Here we propose a scheme that a Laguerre Gaussian laser irradiates a near-critical-density (NCD) plasma to generate a quasi-monoenergetic and low-divergence proton beam. The Laguerre Gaussian laser pulse in NCD plasma excites a moving longitudinal electrostatic field with a large amplitude, and it maintains the inward bowl-shape for dozens of laser durations. This special distribution of the longitudinal electrostatic field can accelerate and converge the protons simultaneously. Our particle-in-cell (PIC) simulation shows that the efficient proton acceleration can be realized with the Laguerre Gaussian laser intensity ranging from $3.9 \times 10^{21}$ W$cm^{-2}$ – $1.6 \times 10^{22}$ W$cm^{-2}$ available in the near future, e.g., a quasi-monoenergetic proton beam with peak energy ~115 MeV and divergence angles less than $5^0$ can be generated by a $5.3 \times 10^{21}$ W$cm^{-2}$ pulse. This work could provide a reference for the high-quality ion beam generation with PW-class laser systems available recently.

## 1. Introduction

With recent rapid development of ultra-short ultra-intense lasers, proton beams

This work was supported by the Strategic Priority Research Program of Chinese Academy of Sciences (Grant No. XDA25050300), National Natural Science Foundation of China (Grant No. 12205366), the National Key R\&D Program of China (Grant No. 2018YFA0404801), and the Fundamental Research Funds for the Central Universities (Grant Nos. 2020MS138) and the Research Funds of Renmin University of China (20XNLG01).
† Corresponding author. E-mail: liumeng@ncepu.edu.cn
†† Corresponding author. E-mail:weiminwang1@ruc.edu.cn

with cut-off energy of near 100 MeV can be generated from the interaction of intense lasers with solid foils[1]. A few features of the laser-driven ion accelerator, including higher ion density and shorter bunch or more compact size[5], can be applied in various prospective applications, covering proton radiography for cancer treatment[7], fast ignition of nuclear fusion[8] and nuclear physics[9], etc. However, a quasi-monoenergetic beam with low divergence is still a critical requirement for most of these attracting applications under the present laboratory conditions. Usually the proton beams have a divergence up to tens of degrees generated in experiments[11].

To overcome this limitation, significant efforts have been devoted to concentrate the transversely distributed proton beam at the center. For instance, curved targets[14], density-modulated targets embedded in a thick enough substrate[15], and plane targets with rectangular and cylindrical geometries at the back[16] have been widely used to control the ion trajectory near the laser propagation axis. Improving the beam focusing properties due to the charge-separated fields on the surface of the structure targets can efficiently guide proton beams. In addition, a new type of movable electrostatic field with a helical wire excited by laser irradiating on a target can simultaneously concentrate and accelerate the protons inside[17]. However, the application of these laser-driven proton beams is restricted by the expense of the complex assembled target. It is highly desired to promote simple methods to generate a quasi-monoenergetic proton beam with low divergence for the coming petawatt(PW) laser systems.

In this paper, we propose to adopt a relativistic Laguerre-Gaussian (LG) laser pulse with an intrinsic hollow intensity distribution to accelerate and focus a proton beam in a near-critical-density (NCD) target. Such a LG laser pulse with the intensity above $10^{19} W cm^{-2}$ has been available [18] and even higher intensity $10^{22} W cm^{-2}$ of the LG laser could be delivered from the SULF 10PW and SEL 100PW laser facilities in the coming few years [19]. With the intensity ranging from $10^{21} W cm^{-2} - 10^{22} W cm^{-2}$, we find that quasi-monoenergetic proton beams can be efficiently generated when a LG laser pulse interacts with a NCD target. The relativistic laser pulse can excite a large-amplitude longitudinal electrostatic field to accelerate the protons forward. Meanwhile, the centripetal component of this longitudinal electrostatic field substantially converges the protons near the laser propagating axis. While a conventional Gaussian laser pulse with the same intensity irradiates on the NCD plasma, the generated proton beam is usually non-monoenergetic in our PIC simulations. Moreover, the peak energy and the corresponding proton yields using LG laser are higher than the case with the Gaussian laser pulse.

2. **Vacuum field structure of a helical laser pulse**

To obtain ultra-relativistic intensity LG laser pulse, a spirally varying thickness for the spiral phase plate should be used [18]. By making a paraxial approximation to the wave equation of LG laser in cylindrical coordinates, one can obtain the Eigen solution of Laguerre Gaussian beam wave equation as follows [20]:

$$u(r,\varphi,x) = \frac{C_l^p}{\omega(x)}\left(\frac{\sqrt{2}r}{\omega(x)}\right)^{|l|} exp\left(-\frac{r^2}{\omega^2(x)}\right) \times L_p^{|l|}(\zeta) exp\left(-ik\frac{r^2}{2R(x)}\right)$$

$$\times exp(-il\varphi) \, exp(-ikx) \, exp(i\Psi(x)) \qquad (1)$$

The laser is incident at the position $x=0$ and propagates along the $+x$ direction, $r$ is the distance from a spatial arbitrary point to the x-axis, $\varphi$ is the rotating angle around the x-axis. $l$ is the azimuthal quantum number, which can take any integer. $p$ is the radial quantum number, which can take any integer greater than or equal to 0. And $p < |l|$ should be required. All $l$ and $p$ together constitute a set of orthogonal complete $LG_{lp}$ basis vectors, which can represent any spatial mode. For example, the vortex laser with ultra-relativistic laser intensity in the experiment mentioned above is obtained by the superposition of several low-order LG lasers in the form of incoherent superposition [21]. $L_p^l(\zeta)$ is the Laguerre polynomial:

$$L_p^l(\zeta) = \sum_{k=0}^{p} \frac{(l+p)!(-\zeta)^k}{(l+k)!k!(p-k)!}, \quad \zeta = \frac{2r^2}{\omega^2}, \quad C_l^p = \sqrt{\frac{2p!}{\pi(p+|l|)!}} \qquad (2)$$

where $C_l^p$ is the normalized constant, $\omega(x)$ is the radius of the beam at z position, $R(x)$ is the curvature radius, $\Psi(x)_{LG}$ is the Gouy phase of LG laser, $\omega_0$ is beam waist size. They have the following relationships with Rayleigh range $x_R = \pi\omega_0^2/\lambda$:

$$\begin{cases} \omega(x) = \omega_0\sqrt{1+\left(\frac{x}{x_R}\right)^2} \\ R(x) = x\left[1+\left(\frac{x_R}{x}\right)^2\right] \\ \Psi(x)_{LG} = (|l| + 2p + 1)arctan\left(\frac{x}{x_R}\right) \end{cases} \qquad (3)$$

and the spot size $\omega_{l,p}(x)$ and far field divergence angle $\theta_{l,p}$ of LG laser are:

$$\begin{cases} \omega_{l,p}(x) = \sqrt{|l| + 2p + 1}\,\omega(x) \\ \theta_{l,p} = \sqrt{|l| + 2p + 1}\,\theta_g \end{cases} \qquad (4)$$

where $\omega(x)$ is the spot size of Laguerre Gaussian laser in TEM$_{00}$ mode, and $\theta_g$ is the far-field divergence angle. According to Eq. (4), as the angular quantum number $|l|$ and radial quantum number $p$ increase, the beams diverge faster and faster as the propagation distance increases. Thus, the high-order Laguerre Gaussian laser losses faster than the low-order LG laser and is also more difficult to produce. So this paper adopted the TEM$_{00}$ mode ($l = 0$, $p = 0$) Gaussian pulse and the low-order LG pulse.

The amplitude distribution of Laguerre Gaussian laser ($a_{LG}$) and Gaussian laser ($a_G$) is:

$$\begin{cases} a_{LG}(t,r,x;l) = a_0 \cdot e^{-\frac{t^2}{\tau^2}} \cdot e^{-\frac{r^2}{\omega_{bnd}^2}} \cdot \left(\frac{\sqrt{2}\cdot r}{\omega_{bnd}}\right)^{|l|} \cdot L_p^l(\zeta) \\ a_G(t,r) = a_0 \cdot e^{-\frac{t^2}{\tau^2}} \cdot e^{-\frac{r^2}{\omega_{bnd}^2}} \end{cases}$$

$$(5)$$

where $a_0$ is the normalized laser amplitude, $T=\lambda/c$ is laser period, $\tau$ is the pulse duration, $r = \sqrt{z^2 + y^2}$ is the distance from an arbitrary point in the space to x-axis, $x_R = \pi\omega_0^2/\lambda$ is the Rayleigh range, $\omega_{bnd} = \omega_0\sqrt{1 + (x_{spot}/x_R)^2}$ is the laser spot size at the $x_{min}$ interface of the simulation domain, $\omega_0$ is the laser focal spot radius, and $x_{spot}$ is the position of laser focus on the x-axis. When Gaussian laser is converted to Laguerre Gaussian laser and their total laser energy remains constant, the peak laser amplitude $a_0$ will still decrease by $1/e$ compared to the original value. Therefore, in our simulations, Laguerre Gaussian laser and Gaussian laser used the same actual laser amplitude $a_0$, which has the following conversion relationship with the input laser amplitude $a_0'$: $a_0 = a_0' \times (1 - 1/e)$, $(1 - 1/e) \approx 0.63$, at which point the two lasers also have similar total energy[22]. The phase of Gaussian laser and LG laser is also different:

$$\begin{cases} phase_{LG1} = 0 + |l| \cdot \varphi + k_0\left(x_{spot} + \frac{y^2}{2R_C}\right) \\ \qquad\qquad -\Psi(x)_{LG} \\ phase_{G1} = 0 + \frac{2\pi}{\lambda} \cdot \frac{y^2}{2R_C} - \Psi(x)_G \\ phase_{LG2} = \frac{\pi}{2} + |l| \cdot \varphi + k_0\left(x_{spot} + \frac{y^2}{2R_C}\right) \\ \qquad\qquad -\Psi(x)_{LG} \\ phase_{G2} = \frac{\pi}{2} + \frac{2\pi}{\lambda} \cdot \frac{y^2}{2R_C} - \Psi(x)_G \end{cases} \quad (6)$$

Equation (6) depicts that the circularly polarized laser has two linearly polarized beams with the phase difference of $\pi$, and LG laser has a helical phase varying with $\varphi$. $\Psi(x)_G = \arctan(x/x_R)$ is the Gouy phase of the Gaussian beam. The electric field distribution of LG laser in cylindrical coordinates is as follows [23]:

$$E_{LG}(r, x; l) = E_0 \cdot e^{-\frac{r^2}{\omega_{bnd}^2}} \cdot \left(\frac{\sqrt{2} \cdot r}{\omega_{bnd}}\right)^{|l|} \cdot L_p^l(\zeta) \qquad (7)$$

$p=0$, $L_0^l(\zeta) = 1$ is independent on $l$. In this case, the transverse component perpendicular to the direction of propagation can be represented as:

$$E_\perp(r,,\varphi,x; l,\sigma) = E_{LG}(r, x; l)[\sin(\eta + l\varphi)\hat{y} - \sigma \cdot \cos(\eta + l\varphi)\hat{z}]/\sqrt{2} \qquad (8)$$

Here $\eta = kx + \frac{kr^2}{2R_C} - \Psi(x)$, and the longitudinal component parallel to the laser propagation direction is:

$$E_\parallel(r,,\varphi,x; l,\sigma) = E_{LG}(r,x; l)\left[\frac{|l| - ls - 2r^2/\omega(x)^2}{kr}\right] \times \cos[\eta + (l - \sigma)\varphi]\hat{x}$$

(9)

In our simulations, a CP laser carries a spin angular momentum of $\sigma\hbar$. Here $\sigma = 1$ and $\sigma = -1$ represent the left-handed circularly polarized laser and the right-handed CP laser, respectively [26]. Besides, each photon in the LG beam carries an orbital angular momentum (OAM) of $l\hbar$ [27]. Thus, the total angular momentum of each photon is $\boldsymbol{J} = \boldsymbol{L} + \boldsymbol{S}$, that is, each photon carries a total angular momentum of $(l - \sigma)\hbar$. In order to guarantee the LG pulse has an inward ponderomotive force and does not transfer its angular momentum to the protons during the interaction between the laser and NCD plasma, so the angular momentum of photons should equal to 0. [28].

With the parameters of $a_0 = 100$, $l = \pm 1, p = 0, \sigma = 1, \omega_0 = 7\mu m$, we perform 2D PIC simulations for the LG laser interacting with NCD plasma density of $n_e=10n_c$. The result indicates that the LG laser of $l = \sigma = 1$ can gain better acceleration, when the laser will only have focusing pondermotive force and will not transfer angular momentum to protons to impede the acceleration. In addition, according to Eq. (9), when $\boldsymbol{J} \neq 0$, the LG beam will generate a changing longitudinal electric field varying with $(l - \sigma)\varphi$, which may hinder the formation of the uniform electron layer during the "hole boring" stage when laser irradiating on the plasma[22]. Thus, a LG laser of $l = 1, p = 0, \sigma = 1$ is applied in our simulations.

## 3. Simulation results

To achieve the efficient ion acceleration using a LG laser pulse with NCD plasma, we have carried out a series of two-dimensional (2D) particle-in-cell (PIC) simulations using the code EPOCH[29]. The circularly polarized (CP) laser pulse has a wavelength $\lambda = 1$ μm. The amplitude profiles of LG and Gaussian laser are determined by Eq. (5). The normalized amplitude of the CP laser pulse $a'_0 \simeq (I_0\lambda^2/2.74 \times 10^{18})^{1/2}$. In typical simulations, a 30 fs laser pulse has the amplitude $a'_0$ from 60 to 120 and the beam waist size $\omega_0 = 3\lambda$. The wavefront of the laser pulse arrives at the vacuum-foil interface $x = 10\lambda$ at $t$=0. The semi-infinite NCD target is composed of protons $H^+$ and electrons $e^-$ with a uniform density profile. The density $n_e$ changes from $6n_c$ to $10n_c$ as the laser intensity increases. The critical density $n_c = m_e\omega^2/4\pi e^2$, $\omega$ is the angular frequency of the laser, $m_e$ is the electron mass. In each simulation, a 45 $\mu m \times 40$ $\mu m$ simulation box moves along the x-axis at the speed of light and is divided into 3600×1600 cells. The NCD plasma is located at $x \geq 10\lambda$ and each cell has 50 macro-particles in the plasma region. Besides, the Laguerre Gaussian laser and Gaussian laser use the same actual laser amplitude $a_0$, which has the following conversion relationship with the input laser amplitude $a'_0$: $a_0 = a'_0 \times (1 - 1/e)$.

Our simulation shows that a quasi-monoenergetic proton beam is efficiently generated when a LG laser interacts with a NCD plasma, and the energy peak of the protons increases as the growth of laser actual amplitude $a_0$. Figure 1(a) indicates that the energy peak $E_{peak}$ increases from 75 MeV to 500 MeV when $a'_0$ varies from 60 to 120 for LG laser pulse, and also the corresponding proton yield in the energy peak increases, as shown by Fig. 1(b). By comparison, there is no quasi-monoenergetic peak generated or much lower peak energy for the case with Gaussian laser amplitude $a'_0 \lesssim 90$. And the proton numbers in energy peak are obviously lower than those in LG laser, as shown by Fig. 1(b). Besides, although the peak energy generated by a Gaussian laser can reach up to 800 MeV, which is higher than the one of LG laser, the protons yield with high energy is always lower. This indicates that quasi-monoenergetic proton beams can be generated by the interaction of LG laser with NCD plasmas in the near future.

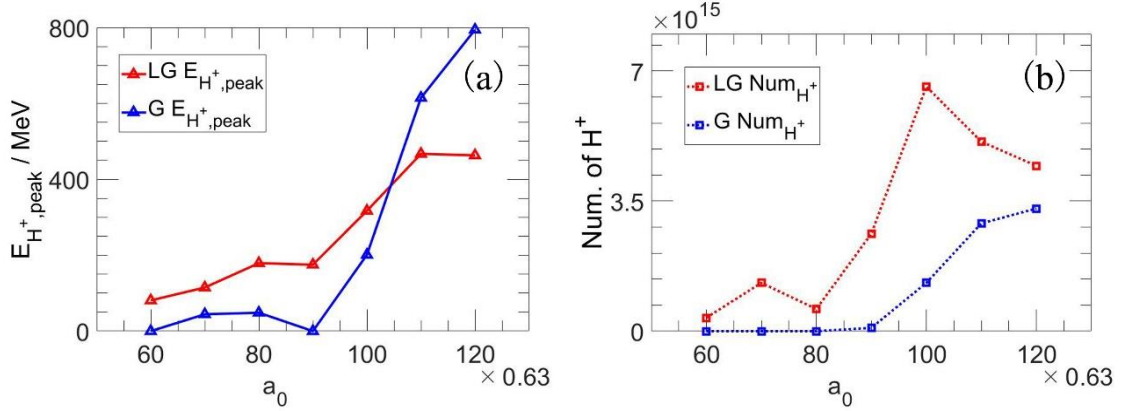

Figure 1: The proton peak energy $E_{H,peak}$ (a) and the corresponding protons yield $Num_H$ in energy peak(b). The red triangle-line and red square dashed-line are the interaction of Laguerre Gaussian with NCD plasma; the blue ones are the cases with Gaussian beam. The laser beam waist size $\omega_0 = 3\lambda$, and the laser amplitude $a_0$ and plasma density $n_e$ are taken as (60, $6n_c$), (70, $7n_c$), (80, $7n_c$), (90, $9n_c$), (100, $10n_c$), (110, $10n_c$), and (120, $10n_c$). Proton filter condition: $|y|<3\ \mu m$.

To further analyze the proton acceleration process, we display the distribution of the longitudinal electrostatic field $E_x$ and the proton density. At the beginning of the LG laser interaction with NCD plasma, as shown in Figs. 2(a) and 2(c), $E_x$ presents three peaks and the two sides are higher. This is because the hollow laser field distribution of LG beam in our simulation provides the ponderomotive force of $F_{y,LG} \propto -\partial|E_\perp|^2/\partial y = 2\partial|E_\perp|^2(2y^2/\omega_0^2 - 1)/y$, which is inward when $y < \omega_0/\sqrt{2}$, while becomes outward when $y > \omega_0/\sqrt{2}$. Not only that, but the laser field is non-zero at center due to the small focal spot, which pushed the electrons forward and meanwhile converged to the x-axis. Consequently, the inducing charge-separated field accelerates and accumulates amounts of protons at $|y| \simeq 1.5\ \lambda$. The distribution of $E_x$ at $t = 45\ T$, as shown by Fig. 2(e), characterized by inward bowl-shape, which significantly accumulates the accelerated protons to the center. In contrast, Gaussian laser excites a charge-separated field with Gaussian-like distribution because the outward pondermotive force of Gaussian laser is $F_{y,G} \propto -\partial|E_\perp|^2/\partial y = 4|E_\perp|^2 y/\omega$. Thus, both the electrons and protons are repelled laterally and there are only a few protons are accelerated forward around the wavefront, as shown by Figs. 2(f) and 2(h).

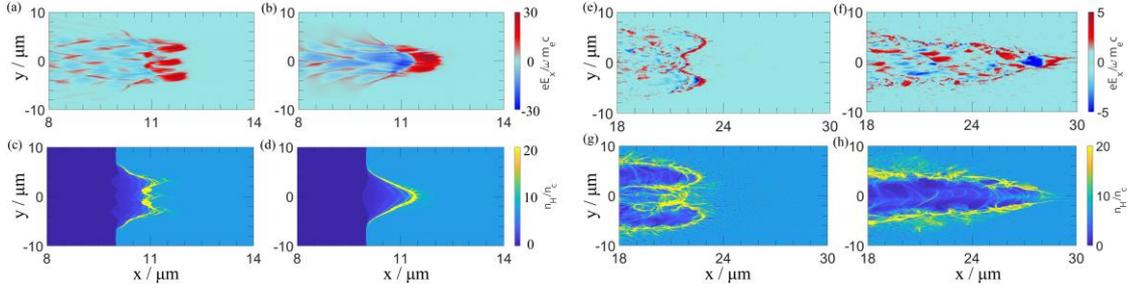

Figure 2: Spatial distribution of the longitudinal electrostaticfield $E_x$ (a,b,e,f) and proton density $n_H / n_c$ (c,d,g,h)when $a_0' = 70$. The comparisions of LG laser and Gaussian laser at 10T are taken as (a, c) and (b, d); the comparisions of LG laser and Gaussian laser at 45T are taken as (e, g) and (f, h). The laser amplitude and target density are taken as (70, $7n_c$).

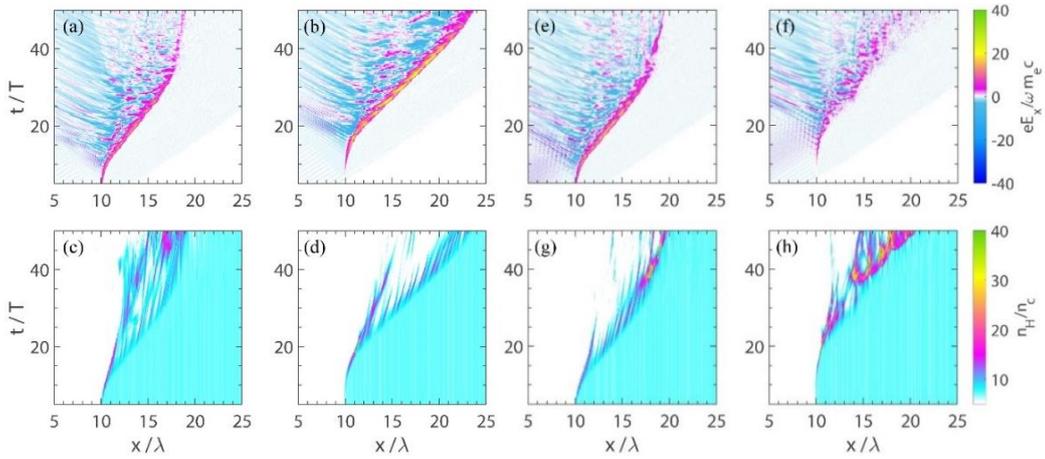

Figure 3: The temporal evolution of longitudinal electrostaticfield distribution $E_x$ (a, b, e, f) and proton density distribution (c, d, g ,h). (a)(c) and (b)(d) present the temporal evolutions at y=0 for LG laser and Gaussian laser. (e)(g) and (f)(h) present the temporal evolutions at y=3 for LG laser and Gaussian laser. The laser plasma parameters are the same as those in Fig. 2.

To characterize the accelerating field excited by LG laser and Gaussian laser in NCD plasma, we compare the temporal evolutions of $E_x$ and proton density $n_H$ in Fig. 3. Figs 3.(a) and 3(e) indicate that the longitudinal electric field $E_x$ excited by Gaussian laser has a greater intensity at $y = 3\lambda$ than the position of $y = 0$, which efficiently accelerates and converges the protons simultaneously, as shown by Fig. 3(c)(g). On the contrary, the longitudinal electric field $E_x$ in Figs. 3(b) and 3(f) shows that $E_x$ excited by Gaussian laser has a greater intensity at $y = 0$, while has a lower intensity at $y = 3\lambda$. As a result, by comparing Figs. 3(d) and 3(h), the protons cannot be accelerated for a long time around the x-axis, but expelled transversely.

In addition, the slope of $E_x$ in Fig. 3 represents the forward moving speed of longitudinal electric field along the x-axis. Combining Figs. 3(a) and 3(e) and Figs. 2 (a) and 2(e), we can indicate that the propagating velocity of $E_x$ at $y = 3\lambda$ is a little faster than the one at $y = 0\lambda$, which maintains the spatial distribution of $E_x$ conducive to proton acceleration. However, $E_x$ of Gaussian laser at y = $0\lambda$ moves faster due to its peak laser intensity. As shown by Figs. 2(h) and 2(f), the longitudinal electrostatic field $E_x$ around x-axis moves so fast that most protons are laterally expelled rather than accelerated forward. Thus, a large amount of protons are accelerated and laterally compressed by the longitudinal electrostatic field $E_x$ excited by LG laser, while only a few protons are accelerated to high energy and diverged by $E_x$ due to the outward ponderomotive force of the Gaussian laser.

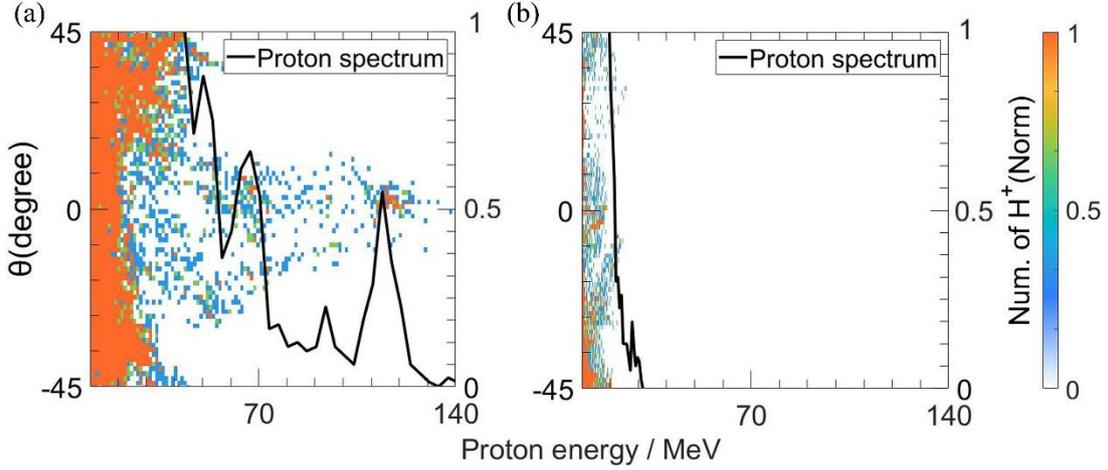

Figure 4: Comparison of the proton energy-angle ($\theta = \arctan(v_y/v_x)$) between LG laser pulse (a) and Gaussian laser pulse (b) at $t = 55T_0$. Proton filter criteria: kinetic energy $E_k >$ 10 MeV and $|y| < 3\ \mu m$. The laser plasma parameters are the same as those in Fig. 2.

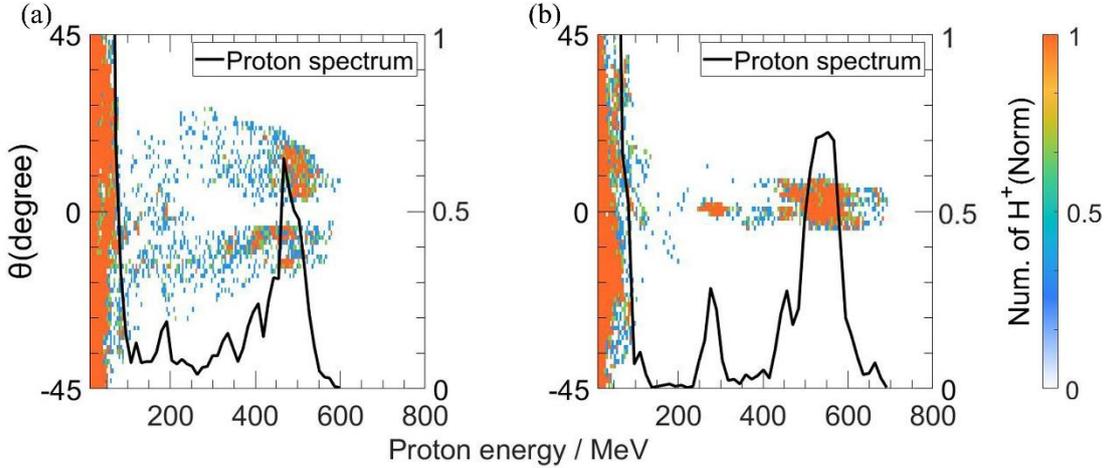

Figure 5: Comparison of the proton energy-angle ($\theta = \arctan(v_y/v_x)$) between LG laser pulse (a) and Gaussian laser pulse (b) at $t = 55T_0$. Proton filter criteria: kinetic energy $E_k >$ 10 MeV and $|y| < 3\ \mu m$. The laser pulses have the focal spot size $\omega_0 = 7\lambda$, and the laser amplitude and target density are taken as (100, $10n_c$).

Consequetly, a energy peak with 115 MeV is generated and the corresponding divergence angle $\theta < 5^0$, as shown in Fig. 4(a). However, due to the repulsion to the protons by the outward pondermotive force of Gaussian laser, as shown in Fig. 4(b), no high-energy proton was observed and the protons are dramatically divergent. Besides, only by increasing the laser intensity to above $10^{22}\ Wcm^{-2}$ and enlarging the laser focal spot to $\omega_0 = 7\lambda$, can we get a quasi-monoenergetic peak for Gaussian laser, as shown in Fig. 5(b). With the laser intensity ranging from $10^{21}$ $Wcm^{-2}$ – $10^{22}$ $Wcm^{-2}$ available in the near future[30], the LG laser pulse with small focal spot can excite a movable longitudinal electric field in the NCD plasma with bowl-shape distribution, achieving better proton acceleration.

4. **Conclusion**

In summary, we have investigated by PIC simulation that a quasi-monoenergetic proton beam can be efficiently generated for a LG laser pulse with a $10^{21} Wcm^{-2}$ – $10^{22} Wcm^{-2}$ in a NCD plasma. With a suitable density range of the NCD plasma, the longitudinal electrostatic field excited can maintain the distribution characterized by the inward bowl-shape for dozens of laser durations. In this way, the protons can be accelerated and converged simultaneously by the electrostatic field, and therefore, a low-divergence proton beam with a peak of 115 MeV was obtained for LG laser with $I_0 \simeq 5.3 \times 10^{21}$ $Wcm^{-2}$. Our simulations have shown that the protons with no or much lower quasi-monoenergetic peak energy are generated by the Gaussian laser with the same intensity. It is worth pointing out that this work could provide a reference for the efficient ion acceleration experiments with PW-class laser systems recently available[30]. Besides, an NCD plasma might be achieved with the help of a secondary nanosecond laser pulse at a lower intensity[32]. Alternatively, NCD plasma might be obtained by the ionization of an ultra low-density plastic foam[33].